\documentstyle[psfig,nicV]{article}
\begin{document}
%
%
\heading{%
The triple--alpha process and its anthropic
significance\\
%
}
\par\medskip\noindent
%
\author{%
Heinz Oberhummer$^{1}$, Rudolf Pichler$^{1}$, Attila 
Cs\'ot\'o$^{2}$
}
\address{
Institute of Nuclear Physics, Vienna University of Technology,
Wiedner Hauptstr.~8--10, A--1040 Wien, Austria 
}
\address{Department of Atomic Physics, E\"otv\"os University, 
Puskin utca 5-7, H--1088 Budapest, Hungary
}
%
\begin{abstract}
Through the triple--alpha process practically all of the carbon
in our universe is synthesized as the ash of helium burning in red giants.
The triple--alpha process proceeds trough the ground state
of $^8$Be and though the 0$^2_+$--state in $^{12}$C. 
We investigate the dependence of 0$^2_+$--state and the production of carbon as a function
of the strength of the underlying nucleon--nucleon interaction. This is
performed by using the complex scaling method in a microscopic cluster model. 
\end{abstract}
\section{Introduction}
The triple--alpha process occurring in helium burning of red giants
is of special significance with respect to the anthropic principle
\cite{CA74,BA86}. The anthropic principle deals with
the question if our universe is tailor--made for the evolution
of life. In other words, could life also have evolved
in the universe, if the
values of the fundamental constants or the initial conditions
in the big bang were different. The reason for the relevance
of the triple--alpha process with respect to the anthropic principle
lies in the fact that one has to deal with physical quantities that
lie in the realm of experimentally verifiable and theoretically calculable
physics. This is for instance hardly the case for the rather uncertain and complicated
science necessary for the description the big bang as well as for the creation
and evolution of life on earth.

The formation of $^{12}$C through hydrogen burning
is blocked by the absence of stable elements for the
mass number $A=5$ and $A=8$. \"Opik and Salpeter pointed out 
\cite{OP51,SA52} that the lifetime of $^8$Be is long enough, 
so that the $\alpha +\alpha \rightleftharpoons\; $$^8$Be 
reaction can produce macroscopic amounts of equilibrium 
$^8$Be in stars. Then, the unstable $^8$Be could capture an 
additional $\alpha$--particle to produce stable $^{12}$C. 
However, this so--called triple--alpha reaction has very 
low rate since the density of $^8$Be in the stellar
plasma is very low because of its short lifetime of 10$^{-16}$\,s. 

Hoyle argued \cite{HO53} that in order to explain the 
measured abundance of carbon in the Universe, the triple--alpha 
reaction cannot produce enough carbon in a non--resonant way, but must 
proceed through a hypothetical resonance of 
$^{12}$C, thus strongly enhancing the cross section. 
Hoyle suggested that this resonance is a $J^\pi=0^+$ state 
at about $\epsilon=0.4$ MeV (throughout this paper $\epsilon$ denotes 
resonance energy in the center-of-mass frame relative to 
the three-alpha threshold, while $\Gamma$ denotes the full 
width). Subsequent experiments indeed found a $0^+$ 
resonance in $^{12}$C in the predicted energy region \cite{HO53,CO57}. It 
is the second $0^+$ state ($0^+_2$) in $^{12}$C. Its modern parameters 
$\epsilon=0.3796$\,MeV and $\Gamma=8.5 \times 10^{-6}$\,MeV 
\cite{AJ88} agree well with the old theoretical prediction.

In the following we discuss in Sect.~2 the used methods,
i.e., the microscopic
three--cluster model, the effective
nucleon--nucleon (NN) interactions, and the complex scaling method.
In Sect.~3 we present the results for the triple--alpha reaction rates using different 
strengths
of the NN--interaction. In Sect.~4 we discuss the astrophysical consequences
of the obtained results.
\section{The model}
Our model is the a microscopic three--cluster ($\alpha$ + $\alpha$ + $\alpha$)
resonating group model approach to the 12--nucleon system. Solving the
12--nucleon Schr\"odinger equation using a three--cluster trial function we
get an equation for the intercluster relative wave function representing
the three--body dynamics of the $^{12}$C states.

In order to avoid any possible
model dependence of the conclusion we use three different effective
NN--interactions: the Minnesota (MN) force designed to reproduce
low--energy scattering data \cite{RE70,TH77}, while the rather different Volkov 1
(V1) and 2 (V2) forces where obtained from fitting the bulk properties
of s-- and p--shell nuclei \cite{VO65}. Each force contains an exchange mixture
parameter, $u$ and $m$, respectively. The parameters were chosen
to reproduce the experimental resonance energy of $\epsilon=0.38$\,MeV for
the $0^+_2$--state in $^{12}$C
(MN: $u = 0.941$; V1: $m = 0.568$; V1: $m= 0.594$).

The three--body resonance energies
for the $0^+_2$--state were determined by using the complex scaling method
(CSM). It reduces the problem of asymptotically divergent resonant
states to that of bound states, and can handle the Coulomb interaction without
any problem.

A more detailed discussion of the model described in this section can be
found in Ref.~\cite{PI97}.
\section{Reaction rates for the triple--alpha process}
In this section we investigate the change of the reaction rate by varying the
strength of all attractive and repulsive terms of the effective NN--potential
through multiplication with a factor $p$. The consequences for triple--alpha
reaction rate will be investigated, if this factor is changed
by a very small amount of the order of 0.1\,\%.

The reaction rate for the triple--alpha process proceeding via the ground state of
$^8$Be and the $0^+_2$--resonance in $^{12}$C is given by \cite{RO88}
\begin{equation}
r_{3\alpha} = 3^{\frac{3}{2}} N_{\alpha}^3
\left(\frac{2 \pi \hbar^2}{M_{\alpha} k_{\rm B} T}\right)^3
\frac{\omega \gamma}{\hbar} \exp \left(-\frac{\epsilon}{k_{\rm B} T}\right) 
\label{e1} ,
\end{equation}
where $M_{\alpha}$ and $N_{\alpha}$ is the mass and the number density of the
$\alpha$--particle, respectively. The temperature of the stellar plasma
is given by $T$. The quantity $\epsilon$ denotes the difference in energy between
the  $0^+_2$--resonance in $^{12}$C and the 3$\alpha$--particle
threshold.
The resonance strength $\omega \gamma$ is given by
\begin{equation}
\omega \gamma = \frac{\Gamma_{\alpha} \Gamma_{\rm rad}}
{\Gamma_{\alpha} + \Gamma_{\rm rad}} \approx \Gamma_{\gamma} .\label{e2}
\end{equation}
The approximation of the above expression for the decay widths of the $0^+_2$--resonance
follows, because for the $\alpha$--width $\Gamma_{\alpha}$, 
radiation width $\Gamma_{\rm rad}$, the electromagnetic decay width $\Gamma_{\gamma}$
to the first excited state of $^{12}$C,
and for the electron--positron pair emission
decay width $\Gamma_{\rm pair}$ into the ground state of $^{12}$C
the following approximations hold:
(i)  $\Gamma_{\alpha} \gg \Gamma_{\rm rad}$ and
(ii) $\Gamma_{\rm rad} = \Gamma_{\gamma} + \Gamma_{\rm pair} \approx 
\Gamma_{\gamma}$.

Therefore, Eq.~(1) can therefore approximated by:
\begin{equation}
r_{3\alpha} \approx 3^{\frac{3}{2}} N_{\alpha}^3
\left(\frac{2 \pi \hbar^2}{M_{\alpha} k_{\rm B} T}\right)^3
\frac{\Gamma_{\gamma}}{\hbar} \exp \left(- \frac{\epsilon}{k_{\rm B} T}\right) 
\label{e3} ,
\end{equation}
The two quantities in Eq.~(3) that change its value by varying the effective
NN--interaction is the energy of
the $0^+_2$--resonance $\epsilon$ in $^{12}$C
and its electromagnetic decay width
$\Gamma_{\gamma}$.
In Table 1 we show the
change of the energy $\epsilon(p)$ of
the  $0^+_2$--resonance with respect to the 3$\alpha$--threshold
in $^{12}$C as a function of the multiplication of the strength
factor $p$ for the three effective NN--interactions MN, V1 and V2. For
no change we obtain again $\epsilon(1) = \epsilon$.
\begin{center}
\begin{tabular}{c c c c}
\multicolumn{4}{l}{{\bf Table 1.} Change of the energy $\epsilon$ of
the $0^+_2$--resonance in $^{12}$C
with}\\
\multicolumn{4}{l}{respect to the 3$\alpha$--threshold as a function of the
strength factor $p$} \\
\hline
\multicolumn{1}{c}{Effective NN--interaction}&
\multicolumn{1}{c}{MN}&
\multicolumn{1}{c}{V1}&
\multicolumn{1}{c}{V2}\\
\hline
\multicolumn{1}{c}{$p$}&
\multicolumn{1}{c}{$\epsilon(p)$ [keV]}&
\multicolumn{1}{c}{$\epsilon(p)$ [keV]}&
\multicolumn{1}{c}{$\epsilon(p)$ [keV]}\\
\hline
1.002 & 327.5  & 337.5 & 343.7\\
1.001 & 353.7  & 358.7 & 361.7\\
1.000 & 379.6  & 379.6 & 379.6\\
0.999 & 405.2  & 400.3 & 397.2\\
0.998 & 430.5  & 420.8 & 414.6\\
\hline
\end{tabular}
\end{center}
It was found that the change of the reaction rate due to the
the enhancement or reduction factor $f_p$ given below
is larger
by between two and three orders of magnitude than due to $\Gamma_{\gamma}$. Therefore,
we neglected the dependence of the reaction rate on $\Gamma_{\gamma}$ by variations
of the effective NN--interaction. The enhancement or reduction for the
triple--alpha reaction rate is then given by
\begin{equation}
f_p = \frac{r_{3\alpha}(p)}{r_{3\alpha}}
\approx \exp\left(\frac{\epsilon-\epsilon(p)}{k_{\rm B}T}\right) . \label{e4}
\end{equation}
In Table 2 the change of the triple--alpha reaction rate at a temperature of
10$^8$\,K given by the factor $f_p$ is 
shown as a function of the multiplication
of the strength factor $p$
for the three effective NN--interactions MN, V1 and V2.
\begin{center}
\begin{tabular}{c c c c}
\multicolumn{4}{l}{{\bf Table 2.} Change of the triple--alpha reaction rate at a}\\
\multicolumn{4}{l}{temperature of 10$^8$\,K as a function of the strength factor $p$} \\
\hline
\multicolumn{1}{c}{Effective NN--interaction}&
\multicolumn{1}{c}{MN}&
\multicolumn{1}{c}{V1}&
\multicolumn{1}{c}{V2}\\
\hline
\multicolumn{1}{c}{$p$}&
\multicolumn{1}{c}{$f_p$}&
\multicolumn{1}{c}{$f_p$}&
\multicolumn{1}{c}{$f_p$}\\
\hline
1.002 & 422   & 132   & 64.4\\
1.001 & 20.2  & 11.4  & 7.9\\
1.000 & 1.0   & 1.0   & 1.0\\
0.999 & 0.05  & 0.09  & 0.13\\
0.998 & 0.003 & 0.008 & 0.02\\
\hline
\end{tabular}
\end{center}

Table 2 shows that the reaction rate $f_p$ at 10$^8$\,K is enhanced or reduced by the
huge amount of about 4 orders
of magnitude compared to the corresponding variations of the effective NN--interaction
factor given by $p$. Furthermore, the model dependence due to the different used
effective NN--interaction for $f_p$ is less than one order of magnitude, and therefore
much less than the before mentioned enhancement or reduction. Tables 1 and 2 also show
at least for the considered small variations of the effective NN--interaction
a linear scaling of $\epsilon$ and therefore an exponential scaling of
$f_p$ with $p$.
\section{Astrophysical consequences}
The significance of low and intermediate and massive stars for the nucleosysnthesis
of carbon is still unclear \cite{GU96}. Some authors claim that AGB stars must be
dominating in the production of carbon (e.g., \cite{SA91}), whereas others favor the
production of carbon in massive stars (e.g., \cite{PR94}). 
In Ref.~\cite{LI89}
the change of core helium burning in a massive star of 20\,$M_{\odot}$ as well
as shell helium burning in a AGB
star of 5\,$M_{\odot}$ was investigated. In this paper only hypothetical {\it ad hoc\/}
shifts of the resonance energy
of the $0_2^+$--state were investigated, whereas in this work we started by
variations of the NN--interaction.

We can apply some of the results of
Ref.~\cite{LI89} to our results. A lowering of the $0_2^+$ resonance energy
by about 60\,keV
corresponding to a 0.2--0.4\,\% strengthening of the nucleon--nucleon interaction
would lead to about a fourfold increase of the carbon production in a 20--$M_{\odot}$
star.  An increase of the $0_2^+$--state by
about 60\,keV corresponding to a 0.2--0.4\,\% weakening of the nucleon--nucleon
interaction would lead to a decrease of roughly a factor two to three of
the $^{12}$C--abundance in a
20\,$M_{\odot}$ star. For a 5\,$M_{\odot}$ star the situation is not so clear, since
the change of carbon production the changes in the strength of the thermal
pulses may compensate this effect. If the level is increased by about 650\,keV
corresponding to a about 2--4\,\% weakening of the NN--interaction (assuming
a linear scaling of the resonance energy with the NN--interaction)
then practically no more carbon could be produced in
core and shell helium burning.
\section{References}
\acknowledgements{This work is dedicated to Michael Benedikt on the occasion of his
retirement from the University of Vienna. We acknowledge support by the Fonds
zur wissenschaftlichen Forschung in \"Osterreich, project P10361--PHY.}

\vfill
\end{document}